\documentclass[12pt]{article}
\usepackage{latexsym}
\usepackage{bm}
\begin{document}
\topskip=1in

\newcommand{\pd}{\partial}
\newcommand{\foot}{\footnotemark\ }
\newcommand{\beq}{\begin{equation}}
\newcommand{\beqa}{\begin{eqnarray}}
\newcommand{\nonum}{\nonumber}
\newcommand{\nndt}{\noindent}
\newcommand{\eeqa}{\end{eqnarray}}
\newcommand{\eeq}{\end{equation}}
\newcommand{\al}{\alpha}
\newcommand{\be}{\beta}
\newcommand{\ga}{\gamma}
\newcommand{\Ga}{\Gamma}
\newcommand{\lm}{\lambda}
\newcommand{\ep}{\epsilon}
\newcommand{\vep}{\varepsilon}
\newcommand{\si}{\sigma}
\newcommand{\Si}{\Sigma}

\title{Tensor-Scalar Torsion}
\author{Charro Gruver\\ Richard Hammond\\
P.F. Kelly\\ Department of Physics\\ North Dakota State
University\\ Fargo, ND 58102-5566\\}
\date{\today}
\maketitle

\begin{abstract}
A theory of gravity with torsion is examined in which the torsion tensor
is
constructed from the exterior derivative of an antisymmetric rank two
potential plus the dual of the gradient of a scalar field. Field
equations
for the theory are derived by demanding that the action be stationary
under
variations with respect to the metric, the antisymmetric potential, and
the
scalar field. A material action is introduced and the equations of
motion
are derived. The correct conservation law for rotational angular
momentum
plus spin is observed to hold in this theory.

\end{abstract}
\newpage
\section{Introduction}

Recently, a theory of gravitation with torsion of the form\footnote{
\small
Here, and in the remainder of this paper, we shall denote partial
differentiation by comma and (anti)symmetrization over multiple indices
by (square brackets) parentheses.}
\beq\label{tordef1}
S_{\mu\nu\si}=\psi_{[\mu\nu,\si]} \, ,
\eeq
where $\psi_{\mu\nu}$ is an antisymmetric torsion potential has been
investigated \cite{spintorfor}. There are several reasons for our
interest
in this approach. The theory has predictive power owing to the fact that
second-order differential equations for $\psi_{\mu\nu}$ are obtained
when
the curvature scalar is used as the Lagrangian \cite{dynamic}. This
theory
makes definite predictions of a long range spin-spin interaction
\cite{upper}, helicity flip interactions \cite{heliflip}, and radiation
of
torsion waves \cite{tp}, although the effects may be smaller than can be
measured with present technology. In this theory, the source for torsion
is
intrinsic spin, and it has been shown that the correct law for
conservation
of rotational angular momentum plus spin is consistently obtained.

One motivation for analysis of gravitational theories with torsion
is the generic appearance of the antisymmetric tensor field
in models which correspond to the low-energy limit of string theory.
The argument that (\ref{tordef1}) may be related to string theory
first requires the expansion of the curvature scalar $R$ for
$U_4$ spacetime\footnote{
\small In a $U_n$ spacetime the metric or
{\it fundamental tensor} is real-valued and symmetric, and the
connection
is metrically compatible. The additional constraint that the connection
be symmetric defines a $V_n$ spacetime. The details are to be found in
\cite{shouten}. Our notation is such that covariant differentiation with
the Christoffel connection formed from the metric,
$\{_{\al\be}^{\;\;\mu}\}$, shall be indicated by a semi-colon, while the
full covariant derivative, using the entire affine connection including
torsion, is expressed by the operator $\nabla_\mu$.}
in terms of the $V_4$ scalar curvature $\ ^oR$ plus torsion terms
\beq
R={}^oR-S^{\al\be\ga}S_{\al\be\ga} \, .
\eeq
This is equivalent to the low energy string theory Lagrangian, for a
constant dilaton \cite{gts}.

Another reassuring feature of this theory is that when matter sources
are
introduced in the usual manner, consistent physical results are obtained
for phenomenological spin sources and for classical Dirac fields which
possess an intrinsic spin \cite{dirac}. However, in solving the torsion
field equations and the Dirac equation under the {\it ansatz}
(\ref{tordef1}), a scalar field had to be introduced as a function of
integration. Denoting this additional field by $\chi$, the solution for
the
torsion $S^{\mu\nu\sigma}$ in the presence of a source field $\Psi$
which
obeys the Dirac equation is
\beq\label{soln}
S^{\mu\nu\si}=\frac{2\pi i\hbar G}{c^3}
\bar{\Psi}\ga^{[\mu}\ga^{\nu}\ga^{\si ]}\Psi+
\ep^{\mu\nu\si\al}\chi_{,\al} \, .
\eeq
Introducing the dual torsion $b_\mu$, defined as
$b_\mu=\ep_{\mu\al\be\ga}S^{\al\be\ga}$, this may be rewritten as
\beq\label{dualsol}
b^{\mu}=-\frac{12\pi\hbar G}{c^3}
\bar{\Psi}\ga^{\mu}\ga_{5}\Psi-6\chi^{,\mu} \, .
\eeq

To see that the presence of the scalar field is vital to the dynamics of
the general solution, consider the following argument. From the {\it
ansatz} (\ref{tordef1}), it follows that $b^{\si}{}_{;\si}=0$ must hold
identically. For the $b^\mu$ in (\ref{dualsol}), this leads to
\beq
\Box \chi \equiv \chi^{,\mu}{}_{;\mu} =\frac{12\pi\hbar G}{c^3}
\big(\bar{\Psi}\ga^{\mu}\ga_{5}\Psi\big)_{;\mu} \, ,
\eeq
the right hand side of which, in general, is not zero. Therefore, the
scalar field is dynamical and may not be consistently chosen to vanish.

In the present paper, we investigate the consequences of naturally
incorporating a scalar field {\it a priori} by modifying
(\ref{tordef1}).
This seems reasonable on two grounds. The first is that, as shown above,
a
scalar field naturally arises in the solutions of this model. The second
is
that the low-energy limit of string theory possesses scalar modes in
addition to the tensor and vector modes. Perhaps one of these modes
contributes to the gravitational sector in the way that we propose.
Thus,
we shall assume from the start that the torsion tensor admits the
decomposition
\beq\label{tordef}
S_{\mu\nu\si}=H_{\mu\nu\si}+\ep_{\mu\nu\si\al}\lm^{,\al} \, ,
\eeq
where $H_{\mu\nu\si}=\psi_{[\mu\nu,\si]}$, and $\lm$ is the scalar
field.

The main result that we obtain is that no harm is done to the 
physically
desirable law of conservation of total angular momentum (rotation plus
spin). Furthermore, it is still the case in our model that spin acts as
the
source for the $H_{\mu\nu\sigma}$ part of the torsion, while the scalar
field does not directly arise from spin.

\section{Field equations}

The action integral is given by
\beq\label{action}
I=\int(\sqrt{-g}R+k{\cal L}_M) \, d^4x \, ,
\eeq
where $\sqrt{-g}R$ and ${\cal L}_M$ are the gravitational and material
Lagrangian densities respectively, and $k=8\pi G/c^4$. The curvature
scalar, $R=g^{\al\be}R_{\al\be}$, includes torsion terms. With torsion
of
the form (\ref{tordef}), $R$ expands as
\beq\label{ricci}
R={}^oR-H^{\al\be\ga}H_{\al\be\ga}+2\ep^{\al\be\ga\si}H_{\be\ga\si}\lm_{,\al}
+6\lm^{,\al}\lm_{,\al} \, .
\eeq
Variations of $I$ with respect to $g_{\mu\nu}$, $\psi_{\mu\nu}$, and
$\lm$,
are assumed to vanish, yielding the classical field equations. With the
following notational conventions for the sources of the metric,
antisymmetric potential, and scalar fields:
\beq\label{lsymdef}
\int \sqrt{-g}T^{(\mu\nu)}\delta g_{\mu\nu} \, d^{4}x\equiv
\int\frac{\delta {\cal L}_M}{\delta g^{\mu\nu}}
\delta g_{\mu\nu} \, d^{4}x \, ,
\eeq
\beq\label{lasymdef}
\int \sqrt{-g}T^{[\mu\nu]}\delta \psi_{\mu\nu} \, d^{4}x\equiv
\int\frac{\delta {\cal L}_M}{\delta \psi^{\mu\nu}}\delta
\psi_{\mu\nu} \, d^{4}x \, ,
\eeq
\beq\label{phi}
\int\sqrt{-g}U\delta\lm  \, d^4x\equiv
\int\frac{\delta {\cal L}_M}{\delta\lm}\delta\lm  \, d^4x \, ,
\eeq
the field equations are as follows:
\beq\label{graveqn}
{}^oG^{\mu\nu} - 3 H^{\mu}{}_{\alpha\beta} H^{\nu\alpha\beta} +
\frac{1}{2}
g^{\mu\nu} H^{\alpha\beta\gamma} H_{\alpha\beta\gamma} + 6\lambda^{,
\mu}
\lambda^{, \nu}
- 3 g^{\mu\nu} \lambda^{, \alpha} \lambda^{, \alpha}
= k T^{(\mu\nu)} \, ,
\eeq
\beq\label{toreqn}
H^{\mu\nu\si}{}_{;\si}=-\frac{k}{2}T^{[\mu\nu]} \, ,
\eeq
\beq\label{scalareqn}
\Box\lm = \frac{U}{12} \, .
\eeq
Note that the use of the symbol $T^{[\mu\nu]}$ to denote the source of
the
antisymmetric field is to allow the {\it formal} introduction of
$T^{\mu\nu} = T^{(\mu\nu)} + T^{[\mu\nu]}$ as the source for the
combined
metric and torsion potential $g^{(\mu\nu)}+\psi^{[\mu\nu]}$. $U$
describes
the phenomenological source for the scalar field and $^oG$ is the $V_4$
Einstein tensor.

With these field equations we shall develop the equations of motion for
a
test body with intrinsic spin moving through a region of spacetime
containing both gravitational and torsion fields. 
Furthermore, we verify
the consistency of the conservation law for rotational plus intrinsic
spin
angular momentum in the case in which the external torques due to
gravity
and torsion vanish.

\section{Equations of motion for a small test body}

In reference \cite{spintorfor} Hammond adapted the method established
by Papapetrou \cite{pap} to develop the equations of motion for a 
small body in a gravitational and torsion field (no scalar field).
This method is used
here to write the equations of motion for a small body in a combined
gravitational and torsion field in which the scalar field is present.

Papapetrou solved for the motion of a spinning test particle in a
background gravitational field through the effective use of an 
iterative multipole-like expansion.
There are systematic and practical advantages to
this approach \cite{pap}.
Spinless and structureless extended particles are
zeroth-order poles, while spinning particles possess at least a
pole--dipole structure. For small bodies it is possible to neglect
higher-order pole terms to a reasonable approximation.

The $U_4$ Bianchi Identity \cite{shouten} holds when torsion is present.

\beq\label{bianchi}
\nabla_{\nu}G^{\mu\nu}=2S^{\mu\al\be}R_{\be\al}
-S_{\al\be\ga}R^{\mu\ga\be\al} \, .
\eeq
Substitution of the field equations and use of $U_4$ geometrical
identities leads to
\beq\label{conserv}
T^{\mu\nu}{}_{;\nu}=\frac{3}{2}H^{\mu}{}_{\al\be}T^{\al\be}
+\frac{1}{6k}\lm^{,\mu}U \, ,
\eeq
where $T^{\mu\nu}=T^{(\mu\nu)}+T^{[\mu\nu]}$, as was discussed above.

Let $\tau^{\mu\nu}=\sqrt{-g}T^{(\mu\nu)}$, $j^{\alpha\beta} =
\frac{k}{K}T^{[\alpha\beta]}$, and $\tilde{j}^{\alpha\beta} = \sqrt{-g}
j^{\alpha\beta}$ so that a tilde denotes density, where $K$ is an
arbitrary
coupling constant for the spin field introduced by Hammond in
\cite{spintorfor}. It then follows that
\beq\label{divtau}
\tau^{\mu\nu}{}_{,\nu}=\sqrt{-g}T^{(\mu\nu)}{}_{;\nu}
-\{_{\al\be}^{\;\;\mu}\}\tau^{\al\be} \, ,
\eeq
and therefore
\beq\label{divtau2}
\tau^{\mu\nu}{}_{,\nu}=\frac{3K}{2k}H^{\mu}{}_{\al\be}\tilde{j}^{\al\be}
+\frac{1}{6k}
\lm^{,\mu}\tilde{U}-\{_{\al\be}^{\;\;\mu}\}\tau^{\al\be} \, .
\eeq
Consider the integral of $\tau^{\mu\nu}{}_{,\nu}$ over a spatial
hypersurface characterized by fixed time. For a bounded source, the
integral has support only over the finite volume of the body, and hence
the
surface terms which arise upon invoking the divergence theorem vanish
({\it
viz.,} $\int\tau^{\mu i}{}_{,i} \, dV=0$, where Latin indices run over
spatial coordinates). Then,
\beq\label{inttau}
\int\tau^{\mu 0}{}_{,0}dV
=\int \left\{
\frac{3K}{2k}H^{\mu}{}_{\al\be}\tilde{j}^{\al\be}+\frac{1}
{6k}\lm^{,\mu}\tilde{U}-\{_{\al\be}^{\;\;\mu}\}\tau^{\al\be}
\right\}dV \, .
\eeq

Let $y^{\al}$ denote the position of the center of mass of the small
test
body, while $\delta x^{\al}$ are the center of mass coordinates of
points
in the body. Thus $x^{\al} = y^{\al}+\delta x^{\al}$ describes the
position
of any point within the body. Expanding $\{_{\al\be}^{\;\;\mu}\}$,
$H^{\mu}{}_{\al\be}$, and $\lm^{,\mu}$ in Taylor series about $y^{\al}$
to
first order in the small quantities $\delta x^{\alpha}$, and
substituting
these into (\ref{inttau}) results in
\beqa\label{inttauexp}
\frac{d}{dt}\int\tau^{\mu 0}dV&=&
\int \Big\{ \frac{3K}{2k} (H^{\mu}{}_{\al\be}+
H^{\mu}{}_{\al\be,\si} \delta x^{\si} ) \tilde{j}^{\al\be}\\
\nonum
&&+\frac{1}{6k}(\lm^{,\mu}+\lm^{,\mu}{}_{,\si} \delta x^{\si}
)\tilde{U}\\
\nonum
&&-(\{_{\al\be}^{\;\;\mu}\} +\{_{\al\be}^{\;\;\mu}\}_{,\si} \delta
x^{\si}
)\tau^{\al\be} \Big\}dV \, .
\eeqa
The time derivative has been pulled out of the integral on the 
left hand side of (\ref{inttauexp}), and on the right hand side 
it is understood that the fields and their derivatives are
evaluated at the position of the center of mass.

Equation (\ref{inttauexp}) is the equation of motion for the body. To
see
this more clearly, we note that $v^\alpha = d y^\alpha \big/ d\tau$ is
the
center of mass velocity of the body ($\tau$ is the invariant proper time
as
measured along the worldline of the center of mass), and define the
following four quantities:
\beq\label{1}
M^{\mu\nu}\equiv\frac{v^0}{c}\int\tau^{\mu\nu}dV \, ,
\eeq
\beq\label{2}
M^{\al\mu\nu}\equiv -\frac{v^0}{c}\int\delta x^{\al}\tau^{\mu\nu}dV \, ,
\eeq
\beq\label{3}
J^{\mu\nu}\equiv \frac{1}{c}\int(\delta x^{\mu}\tau^{\nu 0}-\delta
x^{\nu}\tau^{\mu 0})dV \, ,
\eeq
\beq\label{4}
m^{\al\mu\nu}\equiv \frac{v^0}{c}\int \delta x^{\al}
\tilde{j}^{\mu\nu}dV
\, .
\eeq
In the language used by Papapetrou, $M^{\mu\nu}$ consists of pole terms,
while $M^{\al\mu\nu}$ and $J^{\mu\nu}$ are dipole contributions.
$m^{\al\mu\nu}$ is the lowest-order torsion term in this approximation
scheme, and is also a dipole term. Reexpressing (\ref{inttauexp}) in
terms
of these quantities produces
\beqa\label{eqnsofmot}
\frac{d}{d\tau}\frac{M^{\mu 0}}{v^0}&=&
\frac{3K}{2k}\frac{v^0}{c}H^{\mu}{}_{\al\be}\int
\tilde{j}^{\al\be}dV+\frac{3K}{2k}H^{\mu}{}_{\al\be,\ga}m^{\ga\al\be}\\
\nonum
&&-\{_{\al\be}^{\;\;\mu}\}M^{\al\be}+\{_{\al\be}^{\;\;\mu}\}_{,\ga}M^{\ga\al\be}\\
\nonum
&&+\frac{1}{6}\frac{v^0}{ck}\int\lm^{,\mu}\tilde{U} dV+
\frac{1}{6}\frac{v^0}{ck}\int
\delta x^{\ga}\lm^{,\mu}{}_{,\ga}\tilde{U} dV \, ,
\eeqa
which, if $H^{\mu}{}_{\al\be}=0$ and $\lm=0$, reduces to the familiar
set
of equations for a rotating body in a gravitational field. Equation
(\ref{eqnsofmot}) is interpreted as follows, the term on the left hand
side
is the net force acting on the test body, and the terms on the right
hand
side are the forces due to the spin field and its gradient, the
gravitational field and its gradient, and the scalar field and its
gradient.

The equations for angular momentum and spin are obtained by integrating
the
divergence of the first moment of the symmetric stress-energy-momentum
density,
\beq\label{ident}
(x^{\si}\tau^{\mu\nu}),_{\nu}=\tau^{\mu\si}+
x^{\si}\frac{3K}{2k}H^{\mu}{}_{\al\be}
\tilde{j}^{\al\be}-x^{\si}\{^{\;\;\mu}_{\al\be}\}\tau^{\al\be}+
\frac{1}{12k} x^{\si}\lm^{,\mu}\tilde{U} \, ,
\eeq
over all space. The boundedness of the source yields $\int
(x^{\mu}\tau^{\nu i})_{,i}dV=0$, simplifying the integral of the left
hand
side of (\ref{ident}).
\beqa\label{integral}
\frac{d}{dt}\int (x^{\si}\tau^{\mu 0})dV&=&\int\tau^{\mu\si}dV+
\int x^{\si}\frac{3K}{2k}H^{\mu}{}_{\al\be}\tilde{j}^{\al\be}dV\\
\nonum
&&-\int x^{\si}\{^{\;\;\mu}_{\al\be}\}\tau^{\al\be}dV+
\int\frac{1}{12k} x^{\si}\lm^{,\mu}\tilde{U} dV \, .
\eeqa
Once again, expanding $\{^{\;\;\mu}_{\al\be}\}$,
$H^{\mu}{}_{\al\be}$, and $\lm^{,\mu}$
about $y^{\al}$, and using the quantities defined above, we see that
\beqa\label{M}
M^{\mu\si}&=&\frac{v^{\si}}{v^0}M^{\mu 0}
-\frac{d}{d\tau}\frac{M^{\si\mu 0}}{v^0}-
\frac{3K}{2k}H^{\mu}{}_{\al\be}m^{\si\al\be}\\
\nonum
&&-\{^{\;\;\mu}_{\al\be}\}M^{\si\al\be}-
\frac{v^0}{12kc}\lm^{,\mu}\int\delta x^{\si}\tilde{U} dV \, .
\eeqa
Following upon its construction, $M^{\mu\sigma}$ must be symmetric
in its indices, which requires, after some simplification, that
\beqa\label{torque}
\frac{d}{d\tau}J^{\si\mu}+v^{\si}\frac{M^{\mu 0}}{v^0}
-v^{\mu}\frac{M^{\si 0}}{v^0}&=&
\frac{3K}{k}H^{[\mu}{}_{\al\be}m^{\si]\al\be}
+2\{^{\;\;[\mu}_{\al\be}\}M^{\si]\al\be}\\
\nonum
&&+\frac{v^0}{kc}\lm^{,[\mu}\int\delta x^{\si]}\tilde{U} dV \, .
\eeqa
The right hand side of (\ref{torque}) is the net applied torque on the
body
due to the spin field, the gravitational field, and the scalar field.
The
terms on the left side represent the proper time derivatives of the
rotational angular momentum plus spin, and the orbital angular momentum.

As defined in (\ref{3}), $J^{\si\mu}$ is the total angular momentum. In
Hammond's original model and again in the present case it is composed of
$L^{\si\mu}$, the rotational angular momentum, plus $S^{\si\mu}$,
arising
from intrinsic spin which is not associated with any physical motion of
the body:
\beq\label{J}
J^{\si\mu}\equiv L^{\si\mu}+S^{\si\mu} \, .
\eeq

When the applied net torque from all sources vanishes, (\ref{torque})
may be integrated along the worldline of the test body to yield an
expression
of the law of conservation of angular momentum
\beq\label{result}
L^{\si\mu}+S^{\si\mu}+y^\si p^\mu-p^\si y^\mu=\rm{constant}.
\eeq

\section{Conclusion}

The existence of rather general arguments in favor of the addition of a
scalar field to the torsion sector of Hammond's theory was in no way a
guarantee that such an augmented model would remain self-consistent. In
this paper we have shown that the inclusion of the scalar field in the
specified manner with a classical phenomenological source may be
achieved
without a breakdown in the conservation laws governing the motion of
classical test bodies. This result suggests that the model deserves
further
investigation.

The only serious deficiency that we have noted is that the source
of the scalar field is nearly unconstrained classically -- unlike the
case of the $\psi_{[\mu\nu]}$ torsion potential in Hammond's original
model.
Now that the augmented model has passed its first hurdle, we plan to
incorporate a classical Dirac field as the source of stress-energy and
spin.
It is hoped that this analysis may shed light upon the physical
origin and properties of the additional scalar field.

\section{Acknowledgements}

This research was funded under a Department Of Energy Traineeship, NASA
grant
NAG81007, NSF EPSCoR grant OSR-9452892, and a North Dakota State
University Grant-In-Aid.

\end{document}